\documentclass[aps,prl,reprint,preprintnumbers,amssymb,nofootinbib]{revtex4-1}
\usepackage[utf8]{inputenc}
\usepackage{graphicx,color}

\newcommand{\bC}{\mathbb{C}}

\newcommand{\cH}{\mathcal{H}}

\newcommand{\bc}{\mathcal{B}}
\newcommand{\bcp}{\mathcal{B}^{+}}
\newcommand{\ou}{\mathbf{e}}

\begin{document}

\title{A predictive framework for quantum gravity and black hole to white hole transition}

\author{Robert Oeckl}

\email{robert@matmor.unam.mx}

\affiliation{Centro de Ciencias Matemáticas,
Universidad Nacional Autónoma de México,
C.P.~58190, Morelia, Michoacán, Mexico}

\date{6 April 2018; 14 July 2018 (v2)}

\preprint{UNAM-CCM-2018-1}

\begin{abstract}

The apparent incompatibility between quantum theory and general relativity has long hampered efforts to find a quantum theory of gravity. The recently proposed positive formalism for quantum theory purports to remove this incompatibility. We showcase the power of the positive formalism by applying it to the black hole to white hole transition scenario that has been proposed as a possible effect of quantum gravity. We show how the characteristic observable of this scenario, the bounce time, can be predicted within the positive formalism, while a traditional S-matrix approach fails at this task. Our result also involves a conceptually novel use of positive operator valued measures.

\end{abstract}

\maketitle

\section{Introduction}

Most approaches to quantum gravity rely substantially both on classical general relativity and on quantum theory. Quantum theory as it is usually understood relies on an a priori notion of time that is essential to the consistent interpretation of joint measurements. That is, the knowledge of the temporal order of measurements is a prerequisite for making predictions about their outcomes. On the other hand, in general relativity it is only the outcomes of measurements that reveal the spacetime structure and their temporal order. This incompatibility between core principles of general relativity and quantum theory in its usual form has posed a serious challenge for any attempt at bringing both theories together \cite{Dys:missedops}.

The most common approach to work around this problem has been to consider a situation where the strong gravity regime is confined to a compact spacetime region. Measurements take place only in an asymptotic region where gravity is weak and the metric is held fixed. This restriction appears to be physically well motivated and in close analogy to how measurements are defined in quantum field theory via the S-matrix. There, one assumes that particles can be approximated as free at very early and very late times in Minkowski spacetime, with interactions confined to intermediate times and treated perturbatively. Evidently, the perturbative treatment of the metric itself is more problematic than the perturbative treatment of other fields living on top of a fixed metric. It is well known that a straightforward quantum field theoretic treatment of perturbative general relativity fails due to non-renormalizability \cite{GoSa:qgtwoloops}. The example in this paper sheds further doubt on whether a perturbative approach in the spirit of the S-matrix to a gravitational theory can succeed even in principle.

It has been argued for some time \cite{Oe:catandclock} that the requirement for an absolute notion of time is not inherent to quantum theory, but an artifact of the \emph{standard formulation} of quantum theory, which was conceived in the 1920s to resemble non-relativistic classical mechanics. A suitable, more fundamental framework for formulating quantum theory is now at hand in the form of the \emph{positive formalism} \cite{Oe:dmf,Oe:firstproc,Oe:posfound}. This does not require an a priori notion of time and is fully compatible with the principles of general relativity, thus doing away with the apparent incompatibility.
We demonstrate in this note how the positive formalism is capable of extracting predictions in quantum gravity where the conventional S-matrix approach fails. We focus on the example of a black hole to white hole transition and the associated bounce time.

\section{A simple black hole bounce model}

In general relativity, a black hole forms when sufficient mass density is reached. Spacetime acquires a singularity inside the black hole, signaling a break down of classical general relativity. It is widely believed that such singularities are an artifact of the purely classical treatment of gravity and will not be present in a quantum theory of gravity \cite{MTW:gr}. One proposed mechanism for avoiding singularities is that of a \emph{bounce} \cite{FrVi:shpsymcolqg}. That is, when in-falling matter starts to form a black hole, an effective repulsive force arises from quantum effects. This eventually leads to the formation of a white hole, which is a time reversed black hole, expulsing all matter to the surrounding spacetime. Note that this is distinct from the well established effect of Hawking radiation \cite{Haw:pcreationbh}, which we neglect.

\begin{figure}
  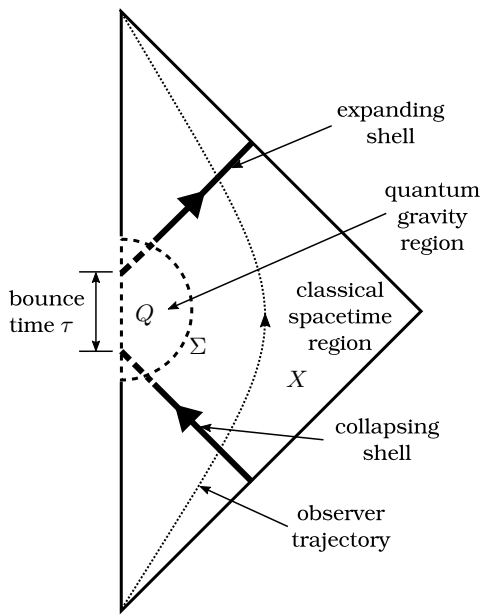
  \caption{\label{fig:bw-penrose} Schematic Penrose diagram of black hole to white hole transition. A distant observer can infer the bounce time $\tau$ by observing the passage, first of the collapsing and then of the expanding shell.}
\end{figure}

To model this process in the simplest possible way, we consider an in-falling spherically symmetric shell of matter with flat Minkowski spacetime in the interior. We further suppose that this shell is infinitesimally thin and contracts at the speed of light. Physically, there is only one relevant parameter that characterizes this contraction process, the mass-energy $m$ of the shell. In classical general relativity a black hole of mass $m$ would form and that would be the end of the story. We assume on the contrary that quantum effects cause the formation of an infinitesimally thin shell of matter of energy $m$ that expands at the speed of light, leaving a flat Minkowski spacetime in the interior. While no metric satisfying Einstein's equations can describe this process in all of spacetime, it turns out that the initial black hole and final white hole metrics can be matched outside a ``small'' spacetime region that encloses the would-be singularity \cite{AmHa:qsupgravcol}. That is, the process conforms to general relativity everywhere, except in this spacetime region which we suppose to be governed by quantum gravity, see Figure~\ref{fig:bw-penrose}. What is more, the freedom in matching is described by a single parameter $\tau$, called the \emph{bounce time} \cite{HaRo:qgbwtunnel}. A distant observer aware of its local spacetime geometry can measure the bounce time $\tau$ by registering the passages of the collapsing and the expanding shells. It corresponds to the time that would have elapsed between the end of the contraction of the collapsing shell to a point and the start of the expansion of the expanding shell from a point.

In order to formalize the problem, we divide spacetime into two regions, $Q$ and $X$. $Q$ is the strong gravity region, enclosed by the dashed line in Figure~\ref{fig:bw-penrose}. $X$ is the weak gravity region outside of the dashed line that covers the remainder of spacetime. We denote by $\Sigma$ the hypersurface that separates the two regions, indicated by the semicircular part of the dashed line. The classical physics in $X$ is described by two parameters, the shell mass $m$ and the bounce time $\tau$. That is, the space of solutions $L_X$ of the equations of motions in $X$ can be written as $L_X=[0,\infty)\times [0,\infty)$. (The bounce time is taken to be bounded by $0$ from below.) For simplicity we take the \emph{phase space} or \emph{space of initial data} $L_{\Sigma}$ at the hypersurface $\Sigma$ to be identical to $L_X$. (Generically one should expect $L_X$ to restrict to a Lagrangian submanifold of $L_{\Sigma}$ on the hypersurface \cite{KiTu:symplectic}.)

Suppose for the moment that rather than a quantum theory of gravity we considered a modified classical theory of gravity that would cause the bounce. This would yield for each shell mass $m$ a bounce time $\tau_c(m)$. More formally, we would have a space $L_Q$ of solutions of the classical equations of motions in $Q$. On the hypersurface $\Sigma$ this would give rise to the subspace of $L_{\Sigma}$ of those initial data that take the form $(m,\tau_c(m))$ for some $m$, which we shall also call $L_Q$. In this way $L_Q$ is a 1-dimensional subspace of the 2-dimensional phase space $L_\Sigma$.

How can we predict the bounce time from a quantum theory of gravity? Suppose we follow the standard formulation of quantum theory and an S-matrix type approach. We should have a Hilbert space $\cH$ of states of our system that describes its degrees of freedom well at least at early and at late times. At intermediate times interactions become important and in our case even metric spacetime itself ceases to exist as a classical entity. We then expect to describe this intermediate regime through an S-matrix $S:\cH\to\cH$. In the present case the initial and final states should describe the collapsing and the expanding shells respectively. However, viewed separately, neither the initial nor the final state carries any information about the bounce time. On the contrary, the states are the same for any bounce time. To obtain the bounce time we need an observer that continuously measures its surrounding spacetime metric, in particular using a clock. What is more, this is not a time measurement on a fixed metric background. Rather, the asymptotic metric is different for each bounce time and it is precisely this difference that encodes the bounce time.

\section{Essentials of the positive formalism}

The shortcomings of the S-matrix in a quantum gravity context provided an important motivation for the \emph{general boundary formulation} as a new approach to the foundations of quantum theory \cite{Oe:catandclock}. The development of this approach \cite{Oe:gbqft}, originally based on topological quantum field theory \cite{Ati:tqft}, has recently lead to the \emph{positive formalism} \cite{Oe:firstproc,Oe:posfound}.\footnote{The formalism used in earlier papers on the general boundary formulation is now called the \emph{amplitude formalism}. Its relation to the \emph{positive formalism} is explained for bosonic theories in \cite{Oe:dmf,Oe:posfound} and for fermionic theories in \cite{Oe:dmf,Oe:locqft}.}
This provides in particular a formulation of quantum theory that implements locality without requiring a metric spacetime background \cite{Oe:dmf}.

In short, the positive formalism is a framework for codifying physical theories by describing the possible \emph{processes} occurring in them and provides a mechanism for predicting the outcomes of these processes. Processes include measurements, observations, interventions, but also ``free evolution''. In the \emph{local} or \emph{spacetime} version of the positive formalism of interest here, spacetime is cut up into \emph{regions} so that a process is taken to occur in each region. The positive formalism then prescribes how these processes are \emph{composed} and how outcomes for the resulting composite process are predicted. Crucially, at this level of description spacetime does not necessarily carry a fixed metric, but may have only a fixed topology.

In order to parametrize the possible interactions or ``signals'' between processes in adjacent spacetime regions, for each interfacing \emph{hypersurface} $\Sigma$ there is a set $\bcp_{\Sigma}$ of (proper) \emph{boundary conditions}. Mathematically, this is the positive cone of a real partially ordered vector space $\bc_{\Sigma}$ (of \emph{generalized boundary conditions}). This comes from the fact that we are allowed to form new boundary conditions by probabilistically combining given ones with positive weights which sum to one. $\bc_{\Sigma}$ also carries a positive-definite inner product. For a region $M$ a boundary condition $b\in\bcp_{\partial M}$ may encode any information about preparation, interaction with an environment in the surrounding spacetime, post-selection etc., in so far as this is relevant for the prediction of outcomes of measurements in $M$. On the other hand, $b$ may equally well be used to encode the effect of the process in $M$ on the physics in the surrounding spacetime.
The partial order in $\bc_{\Sigma}$ plays the role of organizing boundary conditions into a \emph{hierarchy of generality}. That is, if boundary condition $b$ is a special case of or a restriction of boundary condition $c$ this is expressed as $b\le c$. For example, the exterior temperature being in the range $15-20^\circ C$ is more special than being in the range $10-20^\circ C$.
We also assume the existence of a special boundary condition $\ou\in\bcp_{\Sigma}$ known as the \emph{state of maximal uncertainty}. Physically, this encodes a total lack of knowledge about the interaction between adjacent processes.
Thus, any (suitably normalized) boundary condition $b\in\bcp_{\Sigma}$ satisfies $b\le\ou$.

For our present purposes we need only consider one type of process for any region $M$, encoding free evolution, i.e., the absence of any intervention or observation. In the positive formalism, it is encoded in a positive linear \emph{probability map} $A_M$, assigning a non-negative real number $A_M(b)$ to any boundary condition $b\in\bcp_{\partial M}$. (We use here the notation of \cite{Oe:dmf} rather than that of \cite{Oe:posfound}.) This number may be interpreted as a degree of \emph{compatibility} of the boundary condition $b$ with the presence of the spacetime region $M$. E.g.\ in classical field theory it may encode whether there exist solutions of the equations of motion in $M$ that satisfy given conditions $b$ on the boundary of $M$. In order to obtain the \emph{probability} for a certain outcome, we generally need to calculate the quotient between two compatibilities. Thus, say $c\in\bcp_{\partial M}$ is a boundary condition encoding known facts about the surroundings and $b\in\bcp_{\partial M}$ encodes additional restrictions. This implies $b\le c$. Then, the probability $p$ that these additional restrictions are observed is given by the quotient (formula (3) in \cite{Oe:posfound} or formula (8.1) in \cite{Oe:firstproc}),
\begin{equation}
  p=\frac{A_M(b)}{A_M(c)} .
  \label{eq:prob}
\end{equation}
Crucially, $p$ behaves indeed as a probability with the property $0\le p\le 1$. This is guaranteed by the positivity of $A_M$ and the hierarchical relation of the boundary conditions, $0\le b\le c$. These imply $0\le A_M(b) \le A_M(c)$, as required. (The exceptional case $A_M(c)=0$ would signal that $c$ is incompatible with $M$, i.e., unphysical.)

The positive formalism may encode both classical statistical physics and quantum statistical physics. It will be instructive to start with the classical case, applied to the black hole bounce problem. Thus, we suppose again that the bounce arises from a modified classical theory of gravity. Recall Figure~\ref{fig:bw-penrose} with the hypersurface $\Sigma$ separating strong and weak gravity regions. $\bc_{\Sigma}$ is the space of real valued functions on the phase space $L_{\Sigma}$. $\bcp_{\Sigma}$ is the subset of non-negative functions which are interpreted as statistical distributions (without normalization). The probability map $A_Q$ for the strong gravity region $Q$ is given by an integral over $L_Q$ requiring a choice of measure on this space \cite{Oe:posfound}. Taking advantage here of the fact that $L_Q$ may be parametrized by $m$ alone we simply take $\mathrm{d} m$. Thus, for $f\in\bcp_{\Sigma}$,
\begin{equation}
  A_Q(f)  =\int_{0}^\infty f(m,\tau_c(m))\,\mathrm{d}m .
  \label{eq:cprob}
\end{equation}

The question we are interested in might be phrased as follows: What is the probability $p$ that the bounce time lies in the interval $[\tau_1,\tau_2]$ given that the shell mass is $m_0$? The boundary condition $c$ on the phase space imposing the known shell mass is given by the function
\begin{equation}
  c(m,\tau)=\delta(m-m_0) .
  \label{eq:classc}
\end{equation}
The boundary condition $b$ that additionally requires the bounce time to lie in the prescribed interval is,
\begin{equation}
  b(m,\tau)=\delta(m-m_0) \chi_{[\tau_1,\tau_2]}(\tau) .
  \label{eq:classb}
\end{equation}
Here, the characteristic function $\chi_{[\tau_1,\tau_2]}(\tau)$ is one if $\tau_1\le \tau\le\tau_2$ and zero otherwise. Note $b\le c$. As is easy to see, formula (\ref{eq:prob}) yields the expected answer, $p=1$ if $\tau_1\le \tau_c(m_0)\le \tau_2$ and $p=0$ otherwise.

\section{Coherent states and POVM}

While we would like to proceed to a quantum description of the strong gravity region $Q$, the weak gravity region $X$ is well described by classical relativity. \emph{Coherent states} provide a proven tool to interface between a classical and a quantum regime. Say we have a classical phase space $L$ and the Hilbert space $\cH$ of a corresponding quantum system. For each point $\xi\in L$ in phase space we have a corresponding coherent state $|K_{\xi}\rangle\in\cH$ such that the quantum system in this state behaves in a suitable approximation as the classical system in state $\xi$. For the present purposes it is not sufficient, however, to consider individual state space points in isolation. Rather we need a notion of quantum states that correspond to subsets of phase space. This is conveniently accomplished by using the \emph{positive operator valued measure} (POVM) associated to the system of coherent states \cite{Hol:statdectheoquant}.

Thus, we require that the coherent states satisfy a completeness relation with respect to a measure $\mu$ on $L$,
\begin{equation}
  \ou=\int_L |K_{\xi}\rangle \langle K_{\xi}|\,\mathrm{d}\mu(\xi) .
  \label{eq:compl}
\end{equation}
If the coherent states are constructed in the sense of Gilmore and Perelomov \cite{ZFG:coherent,Per:coherent}, such a relation will naturally arise as part of the construction.
Here, $\ou$ denotes the identity operator on $\cH$. It is the state of maximal uncertainty in the quantum theory, also known as the \emph{maximally mixed state}.
For a real valued function $f$ on $L$ define the operator $\hat{f}$ via,
\begin{equation}
  \hat{f}=\int_L |K_{\xi}\rangle f(\xi) \langle K_{\xi}|\,\mathrm{d}\mu(\xi) .
  \label{eq:quantize}
\end{equation}
($f$ is the \emph{covariant symbol} of the operator $\hat{f}$ \cite{Ber:covcont}.)
If $f$ is a function with non-negative values, the operator $\hat{f}$ is positive. In particular, we may interpret $\hat{f}$ as the density matrix that is the quantum analog of the classical statistical distribution given by $f$. In particular, given a subset $H\subseteq L$ the operator $\hat{\chi}_H$ for the characteristic function $\chi_H$ encodes the quantum system being restricted to $H$. Note $\hat{\chi}_H\le \ou$ and $\hat{\chi}_L=\ou$. Also suppose $H_1,\dots,H_n$ are disjoint subsets of $L$ so that their union is all of $L$. Then, $\hat{\chi}_{H_1}+\cdots+\hat{\chi}_{H_n}=\ou$, representing the fact that we can think of the restrictions to the subsets $H_k$ as a mutually exclusive but complete choice in the quantum theory.

\section{Predicting the bounce time}

We switch to a quantum treatment of the region $Q$. Thus, we suppose that we have a quantum theory of gravity formulated within the general boundary formulation \cite{Oe:gbqft} and thus compatible with the positive formalism. This implies that we have a Hilbert space $\cH_{\Sigma}$ of states on the hypersurface $\Sigma$ and a linear \emph{amplitude map} $\rho_Q:\cH_{\Sigma}\to \bC$ encoding the quantum physics in the region $Q$. Moreover, we suppose that we have coherent states $|K_{m,\tau}\rangle\in\cH_{\Sigma}$ corresponding to the elements $(m,\tau)$ of the classical phase space $L_{\Sigma}$. We suppose these satisfy a completeness relation,
\begin{equation}
  \ou=\int_{L_{\Sigma}} |K_{m,\tau}\rangle \langle K_{m,\tau}|\,\alpha(m,\tau)\,\mathrm{d}m\,\mathrm{d}\tau .
  \label{eq:bwcompl}
\end{equation}
The factor $\alpha$ arises due to the fact that the natural measure for the completeness relation ($\mu$ in formula (\ref{eq:compl})) is not in general the measure $\mathrm{d}m\,\mathrm{d}\tau$. Rather, $\mathrm{d}\mu(m,\tau)=\alpha(m,\tau)\,\mathrm{d}m\,\mathrm{d}\tau$. Indeed, due to the non-linear structure of the phase space we should not expect $\alpha$ to be constant.

The space of \emph{boundary conditions} $\bcp_{\Sigma}$ on $\Sigma$ in the positive formalism is here precisely the usual \emph{state space} of statistical quantum theory, up to normalization. That is, $\bc_{\Sigma}$ is the space of self-adjoint operators on $\cH$ with $\bcp_{\Sigma}$ the positive operators. The inner product on this space is the Hilbert-Schmidt inner product, i.e., the trace of the product of operators. The probability map $A_Q$ can be obtained from the amplitude map $\rho_Q$ by making use of the completeness relation (\ref{eq:bwcompl}). For $F\in\bc_{\Sigma}$,
\begin{equation}
  A_Q(F)=\int_{L_{\Sigma}} \rho_Q(F |K_{m,\tau}\rangle) \overline{\rho_Q(|K_{m,\tau}\rangle)}\,\alpha(m,\tau)\,\mathrm{d}m\,\mathrm{d}\tau .
  \label{eq:bwpmap}
\end{equation}
This is equivalent to formula (8) in \cite{Oe:dmf}, but technically more convenient here.

We proceed to set up the boundary conditions in analogy to the classical case. Indeed, we can directly take the classical statistical distributions considered previously and \emph{quantize} them with the POVM via formula (\ref{eq:quantize}).
To this end we may notice that evaluating the probability map (\ref{eq:bwpmap}) on a boundary condition $\hat{a}$ that arises as the quantization of a classical boundary condition $a$ yields the simple expression,
\begin{eqnarray}
  & A_Q(\hat{a})=\int_{L_{\Sigma}} a(m,\tau)\, q(m,\tau)\, \mathrm{d}m\,\mathrm{d}\tau,\qquad\text{with}, \label{eq:qprob} \\
  & q(m,\tau)=|\rho_Q(|K_{m,\tau}\rangle)|^2 \alpha(m,\tau) .
\end{eqnarray}
That is, the predictions of the quantum theory can be captured completely through the statistical distribution $q$ on the classical phase space $L_{\Sigma}$. By comparison, the predictions of a classical theory would be given by a statistical distribution of the form $(m,\tau)\mapsto \delta(\tau-\tau_c(m))$, recovering formula (\ref{eq:cprob}).

Thus, the probability $p$ for the bounce time $\tau$ to lie in the interval $[\tau_1,\tau_2]$ for a shell mass $m_0$ is given by inserting the classical boundary conditions $b$ and $c$ given by (\ref{eq:classc}) and (\ref{eq:classb}) into formula (\ref{eq:qprob}) and taking the quotient (\ref{eq:prob}). This is,
\begin{equation}
  p=\frac{A_Q(\hat{b})}{A_Q(\hat{c})}=\frac{\int_{\tau_1}^{\tau_2} q(m_0,\tau)\, \mathrm{d}\tau}{\int_0^{\infty} q(m_0,\tau)\, \mathrm{d}\tau} .
\end{equation}
Using $q$ we can derive all relevant quantities related to the bounce time. For example its expectation value for shell mass $m_0$ is,
\begin{equation}
  \langle\tau\rangle=\frac{\int_0^{\infty}  \tau\, q(m_0,\tau)\, \mathrm{d}\tau}{\int_0^{\infty} q(m_0,\tau)\, \mathrm{d}\tau} .
\end{equation}

Various elements of the present analysis of the black hole bounce scenario where proposed or treated previously in \cite{CRSV:plancktunnel}, including the introduction of coherent states. Moreover, a concrete proposal was made there for the amplitude $\rho_Q$ from a spin foam model. Based on only partial ingredients of the general boundary formulation, a plausible guess was made in that paper for the formulas for the bounce time. This guess amounted to setting $q(m,\tau)=|\rho_Q(|K_{m,\tau}\rangle)|^2$, missing the crucial factor $\alpha(m,\tau)$. To work out corrected predictions for the bounce time from this spin foam model would require the completeness relation (\ref{eq:bwcompl}), which unfortunately is not provided in that paper.

\section{Conclusion}

In an example scenario we have shown how the positive formalism extends the applicability of quantum theory to realms beyond the reach of the standard formulation and the S-matrix. We propose that it is the proper framework for quantum gravity.

\bigskip

\begin{acknowledgments}
I am indebted to Carlo Rovelli and Simone Speziale for familiarizing me with their work on the bounce scenario and asking me how to correctly calculate the bounce time at a November 2016 meeting in Marseille. The present work elaborates on the answer I gave at that time. This work was partially supported by CONACYT project grant 259258.
\end{acknowledgments}

\bibliography{stdrefsb}

\begin{thebibliography}{20}%
\makeatletter
\providecommand \@ifxundefined [1]{%
 \@ifx{#1\undefined}
}%
\providecommand \@ifnum [1]{%
 \ifnum #1\expandafter \@firstoftwo
 \else \expandafter \@secondoftwo
 \fi
}%
\providecommand \@ifx [1]{%
 \ifx #1\expandafter \@firstoftwo
 \else \expandafter \@secondoftwo
 \fi
}%
\providecommand \natexlab [1]{#1}%
\providecommand \enquote  [1]{``#1''}%
\providecommand \bibnamefont  [1]{#1}%
\providecommand \bibfnamefont [1]{#1}%
\providecommand \citenamefont [1]{#1}%
\providecommand \href@noop [0]{\@secondoftwo}%
\providecommand \href [0]{\begingroup \@sanitize@url \@href}%
\providecommand \@href[1]{\@@startlink{#1}\@@href}%
\providecommand \@@href[1]{\endgroup#1\@@endlink}%
\providecommand \@sanitize@url [0]{\catcode `\\12\catcode `\$12\catcode
  `\&12\catcode `\#12\catcode `\^12\catcode `\_12\catcode `\%12\relax}%
\providecommand \@@startlink[1]{}%
\providecommand \@@endlink[0]{}%
\providecommand \url  [0]{\begingroup\@sanitize@url \@url }%
\providecommand \@url [1]{\endgroup\@href {#1}{\urlprefix }}%
\providecommand \urlprefix  [0]{URL }%
\providecommand \Eprint [0]{\href }%
\providecommand \doibase [0]{http://dx.doi.org/}%
\providecommand \selectlanguage [0]{\@gobble}%
\providecommand \bibinfo  [0]{\@secondoftwo}%
\providecommand \bibfield  [0]{\@secondoftwo}%
\providecommand \translation [1]{[#1]}%
\providecommand \BibitemOpen [0]{}%
\providecommand \bibitemStop [0]{}%
\providecommand \bibitemNoStop [0]{.\EOS\space}%
\providecommand \EOS [0]{\spacefactor3000\relax}%
\providecommand \BibitemShut  [1]{\csname bibitem#1\endcsname}%
\let\auto@bib@innerbib\@empty
\bibitem [{\citenamefont {Dyson}(1972)}]{Dys:missedops}%
  \BibitemOpen
  \bibfield  {author} {\bibinfo {author} {\bibfnamefont {F.~J.}\ \bibnamefont
  {Dyson}},\ }\href {\doibase 10.1090/S0002-9904-1972-12971-9} {\bibfield
  {journal} {\bibinfo  {journal} {Bull. Amer. Math. Soc.}\ }\textbf {\bibinfo
  {volume} {78}},\ \bibinfo {pages} {635} (\bibinfo {year} {1972})}\BibitemShut
  {NoStop}%
\bibitem [{\citenamefont {Goroff}\ and\ \citenamefont
  {Sagnotti}(1985)}]{GoSa:qgtwoloops}%
  \BibitemOpen
  \bibfield  {author} {\bibinfo {author} {\bibfnamefont {M.~H.}\ \bibnamefont
  {Goroff}}\ and\ \bibinfo {author} {\bibfnamefont {A.}~\bibnamefont
  {Sagnotti}},\ }\href {\doibase 10.1016/0370-2693(85)91470-4} {\bibfield
  {journal} {\bibinfo  {journal} {Phys. Lett. B}\ }\textbf {\bibinfo {volume}
  {160}},\ \bibinfo {pages} {81} (\bibinfo {year} {1985})}\BibitemShut
  {NoStop}%
\bibitem [{\citenamefont {Oeckl}(2003)}]{Oe:catandclock}%
  \BibitemOpen
  \bibfield  {author} {\bibinfo {author} {\bibfnamefont {R.}~\bibnamefont
  {Oeckl}},\ }\href {\doibase 10.1088/0264-9381/20/24/009} {\bibfield
  {journal} {\bibinfo  {journal} {Class. Quantum Grav.}\ }\textbf {\bibinfo
  {volume} {20}},\ \bibinfo {pages} {5371} (\bibinfo {year} {2003})},\ \Eprint
  {http://arxiv.org/abs/gr-qc/0306007} {gr-qc/0306007} \BibitemShut {NoStop}%
\bibitem [{\citenamefont {Oeckl}(2013)}]{Oe:dmf}%
  \BibitemOpen
  \bibfield  {author} {\bibinfo {author} {\bibfnamefont {R.}~\bibnamefont
  {Oeckl}},\ }\href {\doibase 10.1007/s10701-013-9741-5} {\bibfield  {journal}
  {\bibinfo  {journal} {Found. Phys.}\ }\textbf {\bibinfo {volume} {43}},\
  \bibinfo {pages} {1206} (\bibinfo {year} {2013})},\ \Eprint
  {http://arxiv.org/abs/1212.5571} {1212.5571} \BibitemShut {NoStop}%
\bibitem [{\citenamefont {Oeckl}(2016)}]{Oe:firstproc}%
  \BibitemOpen
  \bibfield  {author} {\bibinfo {author} {\bibfnamefont {R.}~\bibnamefont
  {Oeckl}},\ }\href@noop {} {\enquote {\bibinfo {title} {A first-principles
  approach to physics based on locality and operationalism},}\ }\bibinfo
  {howpublished} {Frontiers of Fundamental Physics 14 (Marseille, 2014),
  PoS(FFP14)171} (\bibinfo {year} {2016}),\ \Eprint
  {http://arxiv.org/abs/1412.7731} {1412.7731} \BibitemShut {NoStop}%
\bibitem [{\citenamefont {Oeckl}()}]{Oe:posfound}%
  \BibitemOpen
  \bibfield  {author} {\bibinfo {author} {\bibfnamefont {R.}~\bibnamefont
  {Oeckl}},\ }\href@noop {} {\enquote {\bibinfo {title} {A local and
  operational framework for the foundations of physics},}\ }\Eprint
  {http://arxiv.org/abs/1610.09052} {1610.09052} \BibitemShut {NoStop}%
\bibitem [{\citenamefont {Misner}\ \emph {et~al.}(1973)\citenamefont {Misner},
  \citenamefont {Thorne},\ and\ \citenamefont {Wheeler}}]{MTW:gr}%
  \BibitemOpen
  \bibfield  {author} {\bibinfo {author} {\bibfnamefont {C.~W.}\ \bibnamefont
  {Misner}}, \bibinfo {author} {\bibfnamefont {K.~S.}\ \bibnamefont {Thorne}},
  \ and\ \bibinfo {author} {\bibfnamefont {J.~A.}\ \bibnamefont {Wheeler}},\
  }\href@noop {} {\emph {\bibinfo {title} {Gravitation}}}\ (\bibinfo
  {publisher} {W. H. Freeman},\ \bibinfo {address} {San Francisco},\ \bibinfo
  {year} {1973})\BibitemShut {NoStop}%
\bibitem [{\citenamefont {Frolov}\ and\ \citenamefont
  {Vilkovisky}(1981)}]{FrVi:shpsymcolqg}%
  \BibitemOpen
  \bibfield  {author} {\bibinfo {author} {\bibfnamefont {V.~P.}\ \bibnamefont
  {Frolov}}\ and\ \bibinfo {author} {\bibfnamefont {G.~A.}\ \bibnamefont
  {Vilkovisky}},\ }\href {\doibase 10.1016/0370-2693(81)90542-6} {\bibfield
  {journal} {\bibinfo  {journal} {Phys. Lett. B}\ }\textbf {\bibinfo {volume}
  {106}},\ \bibinfo {pages} {307} (\bibinfo {year} {1981})}\BibitemShut
  {NoStop}%
\bibitem [{\citenamefont {Hawking}(1975)}]{Haw:pcreationbh}%
  \BibitemOpen
  \bibfield  {author} {\bibinfo {author} {\bibfnamefont {S.~W.}\ \bibnamefont
  {Hawking}},\ }\href {\doibase 10.1007/BF02345020} {\bibfield  {journal}
  {\bibinfo  {journal} {Commun. Math. Phys.}\ }\textbf {\bibinfo {volume}
  {43}},\ \bibinfo {pages} {199} (\bibinfo {year} {1975})}\BibitemShut
  {NoStop}%
\bibitem [{\citenamefont {Ambrus}\ and\ \citenamefont
  {Hájíček}(2005)}]{AmHa:qsupgravcol}%
  \BibitemOpen
  \bibfield  {author} {\bibinfo {author} {\bibfnamefont {M.}~\bibnamefont
  {Ambrus}}\ and\ \bibinfo {author} {\bibfnamefont {P.}~\bibnamefont
  {Hájíček}},\ }\href {\doibase 10.1103/PhysRevD.72.064025} {\bibfield
  {journal} {\bibinfo  {journal} {Phys. Rev. D}\ }\textbf {\bibinfo {volume}
  {72}},\ \bibinfo {pages} {064025} (\bibinfo {year} {2005})},\ \Eprint
  {http://arxiv.org/abs/gr-qc/0507017} {gr-qc/0507017} \BibitemShut {NoStop}%
\bibitem [{\citenamefont {Haggard}\ and\ \citenamefont
  {Rovelli}(2015)}]{HaRo:qgbwtunnel}%
  \BibitemOpen
  \bibfield  {author} {\bibinfo {author} {\bibfnamefont {H.~M.}\ \bibnamefont
  {Haggard}}\ and\ \bibinfo {author} {\bibfnamefont {C.}~\bibnamefont
  {Rovelli}},\ }\href {\doibase 10.1103/PhysRevD.92.104020} {\bibfield
  {journal} {\bibinfo  {journal} {Phys. Rev. D}\ }\textbf {\bibinfo {volume}
  {92}},\ \bibinfo {pages} {104020} (\bibinfo {year} {2015})},\ \Eprint
  {http://arxiv.org/abs/1407.0989} {1407.0989} \BibitemShut {NoStop}%
\bibitem [{\citenamefont {Kijowski}\ and\ \citenamefont
  {Tulczyjew}(1979)}]{KiTu:symplectic}%
  \BibitemOpen
  \bibfield  {author} {\bibinfo {author} {\bibfnamefont {J.}~\bibnamefont
  {Kijowski}}\ and\ \bibinfo {author} {\bibfnamefont {W.~M.}\ \bibnamefont
  {Tulczyjew}},\ }\href {\doibase 10.1007/3-540-09538-1} {\emph {\bibinfo
  {title} {A Symplectic Framework for Field Theories}}}\ (\bibinfo  {publisher}
  {Springer},\ \bibinfo {address} {Berlin},\ \bibinfo {year}
  {1979})\BibitemShut {NoStop}%
\bibitem [{\citenamefont {Oeckl}(2008)}]{Oe:gbqft}%
  \BibitemOpen
  \bibfield  {author} {\bibinfo {author} {\bibfnamefont {R.}~\bibnamefont
  {Oeckl}},\ }\href {\doibase 10.4310/ATMP.2008.v12.n2.a3} {\bibfield
  {journal} {\bibinfo  {journal} {Adv. Theor. Math. Phys.}\ }\textbf {\bibinfo
  {volume} {12}},\ \bibinfo {pages} {319} (\bibinfo {year} {2008})},\ \Eprint
  {http://arxiv.org/abs/hep-th/0509122} {hep-th/0509122} \BibitemShut {NoStop}%
\bibitem [{\citenamefont {Atiyah}(1988)}]{Ati:tqft}%
  \BibitemOpen
  \bibfield  {author} {\bibinfo {author} {\bibfnamefont {M.}~\bibnamefont
  {Atiyah}},\ }\href@noop {} {\bibfield  {journal} {\bibinfo  {journal} {Inst.
  Hautes \'Etudes Sci. Publ. Math.}\ }\textbf {\bibinfo {volume} {68}},\
  \bibinfo {pages} {175} (\bibinfo {year} {1988})}\BibitemShut {NoStop}%
\bibitem [{\citenamefont {Oeckl}(2017)}]{Oe:locqft}%
  \BibitemOpen
  \bibfield  {author} {\bibinfo {author} {\bibfnamefont {R.}~\bibnamefont
  {Oeckl}},\ }\href {\doibase 10.1007/s40509-016-0086-6} {\bibfield  {journal}
  {\bibinfo  {journal} {Quantum Stud. Math. Found.}\ }\textbf {\bibinfo
  {volume} {4}},\ \bibinfo {pages} {59} (\bibinfo {year} {2017})},\ \Eprint
  {http://arxiv.org/abs/1307.5031} {1307.5031} \BibitemShut {NoStop}%
\bibitem [{\citenamefont {Holevo}(1973)}]{Hol:statdectheoquant}%
  \BibitemOpen
  \bibfield  {author} {\bibinfo {author} {\bibfnamefont {A.~S.}\ \bibnamefont
  {Holevo}},\ }\href {\doibase 10.1016/0047-259X(73)90028-6} {\bibfield
  {journal} {\bibinfo  {journal} {J. Multivariate Anal.}\ }\textbf {\bibinfo
  {volume} {3}},\ \bibinfo {pages} {337} (\bibinfo {year} {1973})}\BibitemShut
  {NoStop}%
\bibitem [{\citenamefont {Zhang}\ \emph {et~al.}(1990)\citenamefont {Zhang},
  \citenamefont {Feng},\ and\ \citenamefont {Gilmore}}]{ZFG:coherent}%
  \BibitemOpen
  \bibfield  {author} {\bibinfo {author} {\bibfnamefont {W.~M.}\ \bibnamefont
  {Zhang}}, \bibinfo {author} {\bibfnamefont {D.~H.}\ \bibnamefont {Feng}}, \
  and\ \bibinfo {author} {\bibfnamefont {R.}~\bibnamefont {Gilmore}},\ }\href
  {\doibase 10.1103/RevModPhys.62.867} {\bibfield  {journal} {\bibinfo
  {journal} {Rev. Modern Phys.}\ }\textbf {\bibinfo {volume} {62}},\ \bibinfo
  {pages} {867–927} (\bibinfo {year} {1990})}\BibitemShut {NoStop}%
\bibitem [{\citenamefont {Perelomov}(1986)}]{Per:coherent}%
  \BibitemOpen
  \bibfield  {author} {\bibinfo {author} {\bibfnamefont {A.}~\bibnamefont
  {Perelomov}},\ }\href@noop {} {\emph {\bibinfo {title} {Generalized Coherent
  States and Their Applications}}}\ (\bibinfo  {publisher} {Springer},\
  \bibinfo {address} {Berlin},\ \bibinfo {year} {1986})\BibitemShut {NoStop}%
\bibitem [{\citenamefont {Berezin}(1972)}]{Ber:covcont}%
  \BibitemOpen
  \bibfield  {author} {\bibinfo {author} {\bibfnamefont {F.~A.}\ \bibnamefont
  {Berezin}},\ }\href@noop {} {\bibfield  {journal} {\bibinfo  {journal} {Math.
  USSR Izvestija}\ }\textbf {\bibinfo {volume} {6}},\ \bibinfo {pages} {1117}
  (\bibinfo {year} {1972})}\BibitemShut {NoStop}%
\bibitem [{\citenamefont {Christodoulou}\ \emph {et~al.}(2016)\citenamefont
  {Christodoulou}, \citenamefont {Rovelli}, \citenamefont {Speziale},\ and\
  \citenamefont {Vilensky}}]{CRSV:plancktunnel}%
  \BibitemOpen
  \bibfield  {author} {\bibinfo {author} {\bibfnamefont {M.}~\bibnamefont
  {Christodoulou}}, \bibinfo {author} {\bibfnamefont {C.}~\bibnamefont
  {Rovelli}}, \bibinfo {author} {\bibfnamefont {S.}~\bibnamefont {Speziale}}, \
  and\ \bibinfo {author} {\bibfnamefont {I.}~\bibnamefont {Vilensky}},\ }\href
  {\doibase 10.1103/PhysRevD.94.084035} {\bibfield  {journal} {\bibinfo
  {journal} {Phys. Rev. D}\ }\textbf {\bibinfo {volume} {94}},\ \bibinfo
  {pages} {084035} (\bibinfo {year} {2016})},\ \Eprint
  {http://arxiv.org/abs/1605.05268} {1605.05268} \BibitemShut {NoStop}%
\end{thebibliography}%
\end{document}